\documentclass[a4paper,11pt]{article}
\pdfoutput=1 % if your are submitting a pdflatex (i.e. if you have
\usepackage{jinstpub} % for details on the use of the package, please
\usepackage{lineno}
\usepackage{graphics}
\usepackage{subfig}
\usepackage{epsfig}
\usepackage{amsfonts}
\usepackage{amsmath}
\usepackage{color}
\usepackage{hyperref}

\usepackage{booktabs}
\usepackage{boldline}
\usepackage{arydshln}

\usepackage{multirow}

 \def\eps{\varepsilon }
 \def\e{{\boldsymbol e}}
 \def\B{{\boldsymbol B}}
 \def\ds{\displaystyle }
\newcommand{\VJ}{\mathbf J}

\newcommand{\VI}{\mathbf I}
\newcommand{\VB}{\mathbf B}

\newcommand{\VX}{\mathbf X}

\newcommand{\Vd}{\mathbf d}
\newcommand{\Ve}{\mathbf e}

\newcommand{\Vh}{\mathbf h}
\newcommand{\Vj}{\mathbf j}

\newcommand{\Vm}{\mathbf m}

\newcommand{\Vu}{\mathbf u}

\newcommand{\Vpsi}{\boldsymbol{\psi}}

\newcommand{\asf}{\mathsf{a}}

\newcommand{\Csf}{\mathsf{c}}
\newcommand{\csf}{\mathsf{c}}

\newcommand{\Jsf}{\mathsf{J}}
\newcommand{\lsf}{\mathsf{\ell}}

\newcommand{\Amat}{{\boldsymbol A}}

\newcommand{\Cmat}{{\boldsymbol C}}

\newcommand{\Hmat}{{\boldsymbol H}}

\newcommand{\Lmat}{{\boldsymbol L}}
\newcommand{\Mmat}{{\boldsymbol M}}

\newcommand{\Rmat}{{\boldsymbol R}}
\newcommand{\Smat}{{\boldsymbol S}}

\newcommand{\Acal}{\mathcal{A}}
\newcommand{\Bcal}{\mathcal{B}}

\graphicspath{{.}{./fig/}} %only for graphics
\makeatletter
\def\input@path{{.}{./fig/}}
\makeatother

\newcommand{\Dp}{\Omega_{\mathrm p}} % plasma domain
\newcommand{\Dci}{\Omega_{\mathrm c_{i}}} % coils domain
\newcommand{\Dps}{\Omega_{\mathrm{ps}}} % passive structures domain
\newcommand{\Dch}{\Omega_{\mathrm L}} % plasma chamber domain

\newcommand{\psiN}{\psi_{\mathrm{N}}}

\title{\boldmath First equilibrium reconstruction for ITER with the code NICE}

\author[1]{B. Faugeras\note{Corresponding author.},}
\author[]{J. Blum,}
\author[]{C. Boulbe}

\affiliation{Universit\'e C\^ote d'Azur, CNRS, Inria, LJAD, Parc Valrose, 06108 Nice Cedex 2, France}

\emailAdd{blaise.faugeras@univ-cotedazur.fr}

\abstract{
In this short paper we present the first application of 
the IMAS compatible code NICE
to equilibrium reconstrution for ITER geometry. 
The inverse problem is formulated as a least square problem and the numerical methods implemented in NICE in order to solve it are presented. 
The results of a numerical experiment are shown: a reference equilibrium is computed from which a set of 
synthetic magnetic measurements are extracted. Then these measurements are used successfully to reconstruct the equilibrium of the plasma.}


\proceeding{International Conference on Fusion Reactor Diagnostics - ICFRD2021\\
  September 6-10, 2021\\
  Varenna, Italy}

\begin{document}
\maketitle
\flushbottom

\section{Introduction}

Numerical reconstruction of the plasma equilibrium in a tokamak is an important and long standing subject 
in fusion plasma science \cite{Shafranov:1971,Zakharov:1973,Luxon:1982,Swain:1982,Lao:1985, Blum:1989}. 
The resolution of this inverse problem consists in 
the computation of the poloidal flux function and of the plasma boundary as well as 
the identification of two non-linear source term functions known as $p'$ and $ff'$ 
in the Grad-Shafranov equation \cite{Grad1958,Shafranov1958,Lust1957}. 
It is needed on the one hand 
for a posteriori analysis of experimental equilibrium configurations 
and on the other hand for real time control of the plasma during a discharge. 
The basic set of measurements needed and used are magnetic probes and flux loops which provide values 
of the poloidal magnetic field and flux at several points surrounding the vacuum 
vessel and the plasma. All free boundary reconstruction codes 
(e.g. \cite{Lao:1990,Blum:1990, McCarthy:1999,Zwingmann2003,
ACTI.B.Faugeras.08.1,
Blum2012,ACL.B.Faugeras.14.2,ACL.B.Faugeras.20.2,Moret2015}) primarily use 
these magnetic measurements which proved to be sufficient to identify correctly 
the plasma boundary and the averaged 
plasma current density profile \cite{Blum2012}. 
The goal of this paper is to present a first test of the adaptation of the numerical 
tools developed by the authors, namely the code NICE \cite{ACL.B.Faugeras.20.2}, to the foreseen ITER configuration. 
NICE stands for "Newton direct and Inverse Computation for Equilibrium".

Next Section \ref{sec:ip-formulation} is devoted to the formulation of the direct model and the inverse problem. 
In Section \ref{sec:ip-discretization} we discuss the numerical methods which 
we have developed for their resolution 
and finally in Section \ref{sec:num} a first test numerical experiment is presented for ITER configuration. 
%on two different test cases.

\section{Inverse problem formulation}
\label{sec:ip-formulation}
\subsection{Free-boundary plasma equilibrium}

The equations which govern the equilibrium of a plasma in the presence of a magnetic field in a 
tokamak are on the one hand Maxwell's equations satisfied in the whole of space (including the plasma):
\begin{equation}
\label{eqn:maxwell}
\begin{aligned}
\nabla \cdot \VB = 0,\ \ & 
\nabla \times (\displaystyle \frac{\VB}{\mu}) =  \Vj, 
\end{aligned}
\end{equation}
and on the other hand the equilibrium equation for the plasma itself
\begin{equation}
\label{eqn:equilibrium}
\nabla p=\Vj \times \VB,
\end{equation}
where $\VB$ is the magnetic field, $\mu$ is the magnetic permeability, $p$ is the kinetic pressure and $\Vj$ is the current density. 
%Equation (\ref{eqn:equilibrium}) implies that for a plasma in equilibrium the field lines 
%and the current lines lie on  isobaric surfaces. These surfaces, generated by the field lines, 
%are called magnetic surfaces. In order for them to remain within a bounded volume 
%of space it is necessary that they have a toroidal topology. These surfaces form a family 
%of nested tori. The innermost torus degenerates into a curve which is called the magnetic axis.
We refer to standard text books 
(e.g. \cite{Freidberg1987, Blum:1989, Wesson2004, Goedbloed2004, Jardin2010}) and to \cite{Heumann2015}
for details of the derivation and only state the needed equations in what follows 
which is a summary of what can be found in \cite{ACL.B.Faugeras.20.2}.

Introducing a cylindrical coordinate system $(\e_r,\e_\phi,\e_z)$ 
($r=0$ is the major axis of the tokamak torus) and assuming axial symmetry 
equations (\ref{eqn:maxwell}) and (\ref{eqn:equilibrium}) reduce to the following equation for 
the magnetic poloidal flux $\psi(r,z)$ in the poloidal plane $\Omega_{\infty}=(0,\infty) \times (-\infty,\infty)$:
\begin{equation}
	- \Delta^* \psi = j_\phi,
	\label{eqn:base}
\end{equation}
where $j_{\phi}$ is the toroidal component of $\Vj$, 
and the second order elliptic differential operator $\Delta^*$ is defined by
\begin{equation}
\label{eqn:delta*}
\Delta^* \psi :=  \partial_r\left(\frac{1}{\mu_0 r}  \partial_r \psi \right) 
+ \partial_z\left(\frac{1}{\mu_0 r}  \partial_z \psi \right)  := \nabla \cdot \left(  \frac{1}{\mu_0 r}  \nabla \psi \right).
\end{equation}
Here $\nabla$ is the 2D operator in the $(r,z)$-plane and $\mu_0$ is the magnetic permeability of vacuum 
(in this work we consider only air-transformer tokamaks such as ITER).

The magnetic field can be decomposed in poloidal and toroidal components 
\begin{equation}
\begin{aligned}
\VB=\VB_p +\VB_{\phi},\ \ &  \VB_p =\ds \frac{1}{r}[\nabla \psi \times \e_{\phi}], \ &
\B_{\phi} = B_\phi \Ve_\phi= \ds \frac{f}{r} \e_{\phi},
\label{eqn:B}
\end{aligned}
\end{equation}
where $f$ is the diamagnetic function. Equation (\ref{eqn:B}) shows that 
%in an axisymmetric configuration 
the magnetic surfaces are generated by the rotation 
of the iso-flux lines around the axis 
%$r=0$ 
of the torus.

The toroidal component of the current density $j_\phi$ 
is zero everywhere outside the plasma domain and the poloidal field coils (and possibly the passive structures). 
The different sub-domains of the poloidal plane of a schematic tokamak 
(see Fig. \ref{fig:tokamak_geometry}) as well as the corresponding expression 
for $j_\phi$ are described below:

\begin{figure}
\subfloat[]{
\centering
\def\svgwidth{0.4\linewidth}
\begingroup%
  \makeatletter%
  \providecommand\color[2][]{%
    \errmessage{(Inkscape) Color is used for the text in Inkscape, but the package 'color.sty' is not loaded}%
    \renewcommand\color[2][]{}%
  }%
  \providecommand\transparent[1]{%
    \errmessage{(Inkscape) Transparency is used (non-zero) for the text in Inkscape, but the package 'transparent.sty' is not loaded}%
    \renewcommand\transparent[1]{}%
  }%
  \providecommand\rotatebox[2]{#2}%
  \ifx\svgwidth\undefined%
    \setlength{\unitlength}{384.59851074bp}%
    \ifx\svgscale\undefined%
      \relax%
    \else%
      \setlength{\unitlength}{\unitlength * \real{\svgscale}}%
    \fi%
  \else%
    \setlength{\unitlength}{\svgwidth}%
  \fi%
  \global\let\svgwidth\undefined%
  \global\let\svgscale\undefined%
  \makeatother%
  \begin{picture}(1,1.14113801)%
    \put(0,0){\includegraphics[width=\unitlength,page=1]{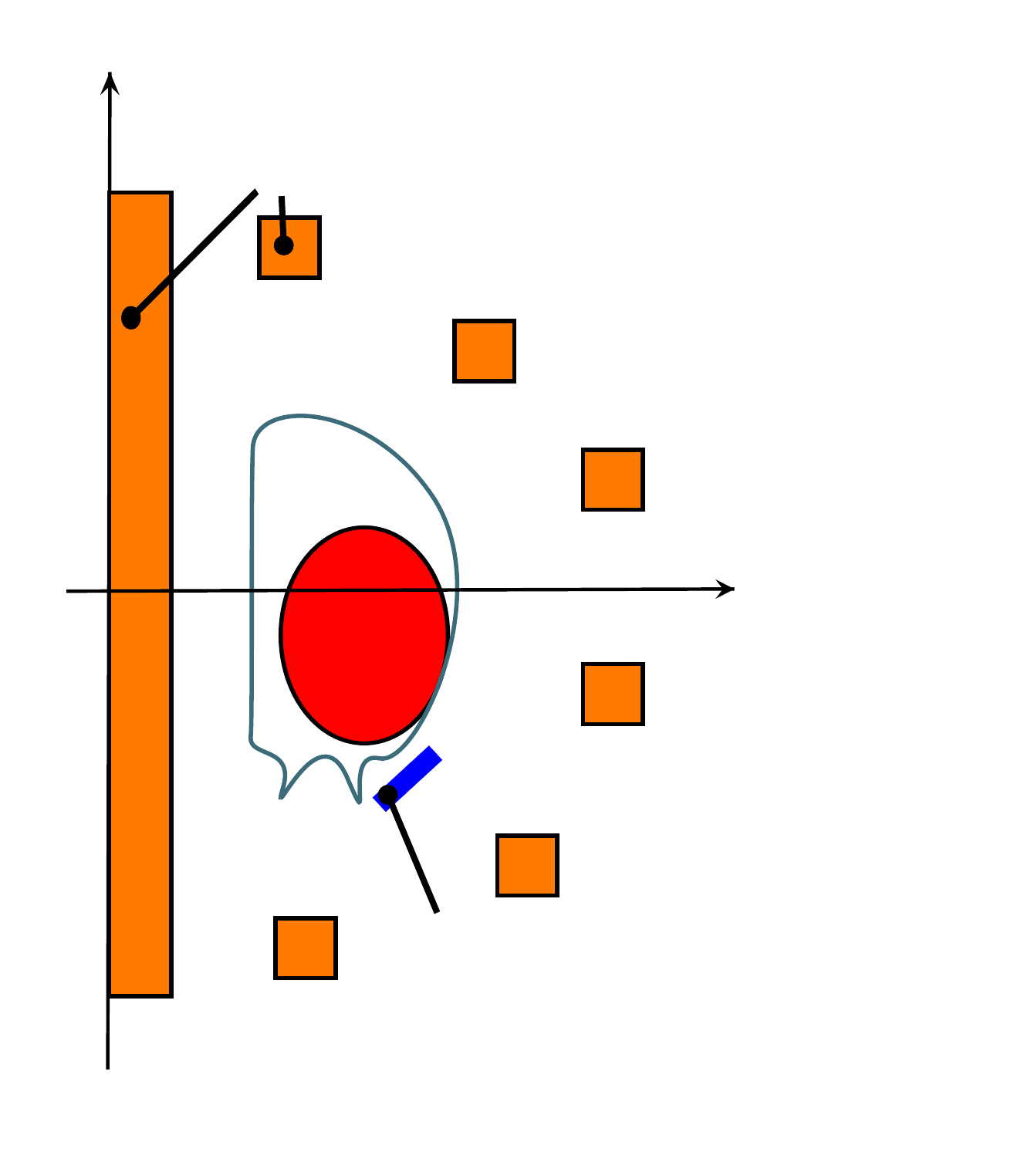}}%
    \put(0.67842924,0.59856416){\color[rgb]{0,0,0}\makebox(0,0)[lb]{\smash{\text{r}}}}%
    %\put(0.00881314,1.04119313){\color[rgb]{0,0,0}\makebox(0,0)[lb]{\smash{\text{z}}}}%
    \put(0.05,1.04119313){\color[rgb]{0,0,0}\makebox(0,0)[lb]{\smash{\text{z}}}}%
    \put(0.22467153,0.94887989){\color[rgb]{0,0,0}\makebox(0,0)[lb]{\smash{ $\Dci$}}}%
    \put(0.38553772,0.18260907){\color[rgb]{0,0,0}\makebox(0,0)[lb]{\smash{ $\Dps$}}}%
    \put(0.27168894,0.48247402){\color[rgb]{0,0,0}\makebox(0,0)[lb]{\smash{ $\Dp$}}}%
    \put(0.22720063,0.65764782){\color[rgb]{0,0,0}\makebox(0,0)[lb]{\smash{ $\Dch$}}}%
    \put(0.34926775,0.86445642){\color[rgb]{0,0,0}\makebox(0,0)[lb]{\smash{ $\partial \Dch$}}}%
    \put(0,0){\includegraphics[width=\unitlength,page=2]{tokamak_geometry_noiron.pdf}}%
    %\put(-0.03166539,0.5398471){\color[rgb]{0,0,0}\makebox(0,0)[lb]{\smash{ $0$}}}%
    \put(0.01,0.5398471){\color[rgb]{0,0,0}\makebox(0,0)[lb]{\smash{ $0$}}}%
  \end{picture}%
\endgroup%
}
\subfloat[]{
\includegraphics[width=0.4\textwidth]{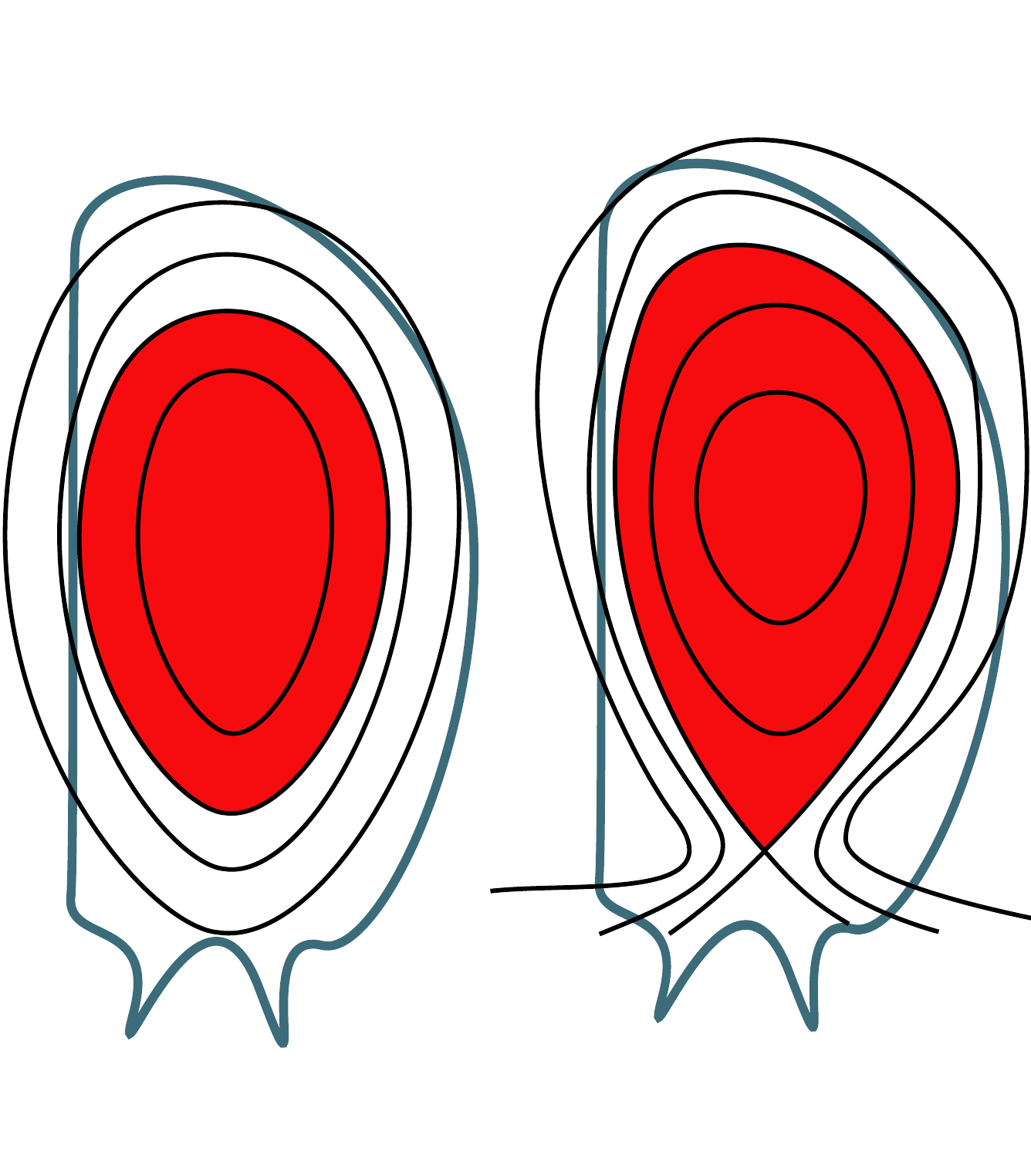}
}
\caption{Left (a): schematic representation of the poloidal plane of a tokamak. $\Dp$ is the plasma domain, 
	$\Dch$ is the limiter domain accessible to the plasma, $\Dci$ represent poloidal field coils and the central solenoid (corresponding 
to the IMAS pf$\_$active IDS), $\Dps$ the passive structures. 
Right (b): example of a plasma whose boundary is defined by the contact with limiter (left) or by the presence of an X-point (right).}
\label{fig:tokamak_geometry}
\end{figure}

\begin{itemize}
	\item[$\bullet$] $\Dch$ is the domain accessible to the plasma. Its boundary is the limiter $\partial \Dch$.

	\item[$\bullet$] $\Dp$ is the plasma domain where equations (\ref{eqn:equilibrium}) and 
		(\ref{eqn:maxwell}) imply that $p$ and $f$ are constant on each magnetic surface i.e. 
		$p=p(\psi)$ and $f=f(\psi)$. One then deduces the so-called Grad-Shafranov equilibrium 
		equation in the plasma \cite{Grad1958,Shafranov1958,Lust1957}
		\begin{equation}
		\label{eqn:gradshaf}
		-\Delta^* \psi = r p'(\psi) + \frac{1}{\mu_0 r}(ff')(\psi).
		\end{equation}
		The right-hand side of (\ref{eqn:gradshaf}) is the toroidal component $j_\phi$ of the current density in the plasma. 

		The plasma domain is unknown, i.e. $\Dp=\Dp(\psi)$, and this is a free boundary problem. 
		This domain is defined by its boundary
		which is the outermost closed $\psi$ iso-contour contained within the limiter $\Dch$. 
		The plasma can either be limited if this iso-contour is tangent to the limiter $\partial \Dch$ 
		or defined by the presence of an X-point (see Fig. \ref{fig:tokamak_geometry}). Functions $p'$ and $ff'$ are zero outside $\Dp$.
		%In the latter  case the plasma domain is strictly bounded by the magnetic separatrix.

		%More precisely
		%\begin{equation*}
		%\begin{array}{c}
		%\Dp(\psi) = \{ (r,z) \in \Dch\ ,  \\ 
		%  \psi(r,z) \geq 
		%\max\left(\underset{(r,z) \in \partial \Dch}{\max}\psi(r,z),
		%	\underset{(r_X,z_X) \in \Dch}{\max} \psi(r_\mathrm{X},z_\mathrm{X}) \right) \},\, 
		%\end{array}
		%\end{equation*}

		%where $(r_\mathrm{X},z_\mathrm{X})$ denotes the coordinates of the saddle points of $\psi$. 

		The current density is non-linear in $\psi$ due to the non-linear functions  $p'$ and
		$ff'$ and the definition of the plasma domain $\Dp(\psi)$.
		While $\Dp(\psi)$ is fully determined for a given $\psi$, the two functions
		$p'$ and $ff'$ are not determined in this modelization. 
		It is the goal of the inverse equilibrium reconstruction problem to determine them.
		For now let us consider that we are given two functions $\Acal(\psiN)$ and $\Bcal(\psiN)$ 
		such that 
		\begin{equation}
		j_\phi=\lambda ( \ds \frac{r}{r_0} \Acal(\psiN) + \ds \frac{r_0}{r} \Bcal(\psiN)).
		\label{eqn:density}
		\end{equation} 
		Here $r_0$ is the major radius of the tokamak vacuum chamber 
		and $\lambda$ is a scaling coefficient. The normalized poloidal flux $\psi_\mathrm{N}(r,z)$ is
		\begin{equation}
		  \psi_\mathrm{N}(r,z) = \frac{\psi(r,z) -
		\psi_{\mathrm{a}}(\psi)}{\psi_{\mathrm{b}}(\psi) - \psi_{\mathrm{a}}(\psi)}\,.
		\label{eqn:psin}
		\end{equation}
		with $\psi_{\mathrm{a}}$ and $\psi_{\mathrm{b}}$ 
		being the flux values at the magnetic axis and at the boundary of the plasma.

		%\begin{equation}
		%\begin{aligned}
		%\psi_{\mathrm{a}}(\psi) &:=\psi(r_{\mathrm{a}}(\psi),z_{\mathrm{a}}(\psi)), \\
		%\psi_{\mathrm{b}}(\psi) &:= \psi(r_{\mathrm{b}}(\psi),z_{\mathrm{b}}(\psi)),   
		%\end{aligned}
		%\label{eq:coordax}
		%\end{equation}

		%with $(r_{\mathrm{a}}(\psi),z_{\mathrm{a}}(\psi))$ the magnetic axis, 
		%where $\psi$ has an extremum in $\Dp(\psi)$ and  
		%$(r_{\mathrm{b}}(\psi),z_{\mathrm{b}}(\psi))$ the coordinates of the point that determines 
		%the plasma boundary. The point $(r_{\mathrm{b}},z_{\mathrm{b}})$ is either an X-point 
		%of $\psi$ or the contact point with the limiter $\partial \Dch$.

		\item[$\bullet$] Each domain $\Dci$, $i=1,\dots N_C$ represent one of the $N_C$ coils carrying currents 
that is to say poloidal field coils as well as  the central solenoid (this corresponds to the IMAS pf$\_$active IDS). The expression of the current density in the $i$-th coil is 
		\begin{equation}
		j_\phi=\ds \frac{I_i}{S_i},
		\label{eqn:jphicoil}
		\end{equation}
		where $S_i$ is the section area of the coil and $I_i$ is a given measured current. 	
		
		\item[$\bullet$] $\Dps$ represents passive structures 
		where the induced current density is assumed to be $0$ in this work 
		but can be considered to be measured 
		and given in the same form as Eq. (\ref{eqn:jphicoil})   
\end{itemize}

To sum up, given functions $\Acal$ and $\Bcal$, and currents $\VI = \{I_{i}\}_{i=1}^{N_C}$ in the coils, 
the free-boundary equilibrium equation for $\psi(r,z)$ on $\Omega_\infty$ 
is the following non-linear boundary value problem
\begin{equation}
\left\{
 \begin{array}{l}
    -\Delta^* \psi  = 
 \begin{cases}
	\lambda ( \ds \frac{r}{r_0} \Acal(\psiN) + \ds \frac{r_0}{r} \Bcal(\psiN)) & \text{in } \Dp(\psi),\,  \\
	\ds \frac{I_i}{S_i} & \text{in } \Dci, \, \\
 	0 & \text{elsewhere},
 \end{cases}\\
    \psi(0,z,t) = 0, \, \\
    \lim\limits_{\|(r,z)\|\to +\infty} \psi(r,z)= 0.\, \\
 \end{array}
\right.
\label{eqn:EquiModel}
\end{equation}

This formulation on an infinite domain is not used directly in computations where we use 
finite elements on a truncated bounded domain. 
The infinite domain is reduced to a semi circular computational 
domain by an uncoupling procedure \cite{Albanese1986,Gatica1995}. 
We chose a semi-circle $\Gamma$ {of radius $\rho_\Gamma$} surrounding 
the coil domains $ \Dci$
% (see Fig. \ref{fig:tokamak_geometry})
and define the computation domain $\Omega$ 
having boundary $\partial \Omega  = \Gamma \cup \Gamma_{0}$, 
where 
%$\Gamma_{0}:=\{(0,z),\, -\rho_\Gamma \le z \le \rho_\Gamma \}$.
$\Gamma_{0}=\{(0,z),\, z\in [-\rho_\Gamma, \rho_\Gamma ] \}$.

The weak formulation of the equilibrium problem on which the finite element method relies uses a 
function space $V$ defined in \cite{Heumann2015} and can be written as:\\
%The weak formulation of the equilibrium problem on which the finite element method relies 
%uses the following function space
%\begin{equation*}
%V:= \left\{\psi : \Omega \to \mathbb R, \|\psi\|
%<
%\infty, \| \frac{|\nabla \psi|}{r}\| < \infty, \, \psi_{|\Gamma_0} = 0 \right\}  \cap C^0(\overline \Omega),
%\end{equation*}
%with
%\[
%\| \psi\|^2  = \int_{\Omega} \psi^2 \, r\, dr dz,
%\]
%and can be written as:\\
Given function $\Acal$ and $\Bcal$, and currents $\VI$, 
find $\psi \in V $ such that for all $\xi \in V$ 
\begin{equation}  
\asf(\psi,\xi) + \csf(\psi,\xi) - \,\Jsf_{\mathrm p}(\psi,\xi;\Acal,\Bcal) - \lsf(\VI,\xi)=0,
\label{eqn:var}
\end{equation}
where 
  \begin{equation}
  \begin{aligned}
  \asf(\psi,\xi) &:= \int_{\Omega} \frac{1}{\mu_0 r} \nabla \psi \cdot
\nabla  \xi \, dr dz,\\
  \Jsf_{\mathrm p}(\psi ,\xi;\Acal,\Bcal) &:=
  \int_{\Dp(\psi)}
 \lambda \left ( \ds \frac{r}{r_0} \Acal(\psiN) 
+ \frac{r_0}{r} \Bcal(\psiN)  \right) \xi \, dr dz,\\
  \lsf(\VI,\xi)&:= \sum_{i=1}^{N_C} \ds \frac{I_i}{S_i}  \int_{\Dci}  \xi \, dr dz, \\
\end{aligned}
 %\label{eq:Asf}
\end{equation}
and the bilinear form $\Csf : V \times V \to \mathbb R$ 
is accounting for the boundary conditions at infinity. 
We refer to \cite{Heumann2015} for its precise expression and to \cite[Chapter 2.4]{Grandgirard1999} for the details 
on its the derivation. Alternative approaches for the incorporation of boundary conditions at infinity 
were more recently presented in \cite{ACL.B.Faugeras.17.3}.

\subsection{The inverse reconstruction problem}

Magnetics constitute the basic set of experimental measurements used in equilibrium reconstruction 
for the identification of functions $\Acal$ and $\Bcal$. 
They consist in measurements of projections 
of the poloidal magnetic field, $ \VB_p \cdot \Vd $ 
at several locations around the vacuum vessel 
of the tokamak (the unit vector $\Vd$ varies with each B-probe) 
and of flux loops measurements, noted $F(\psi)$, at several locations too. 
%(see Fig. \ref{fig:mesh}).

%In order to be able to use polarimetric measurements the electron density function, 
%$N_e(\psiN)$ has to be known. It is therefore also going to be identified using 
%interferometric measurements which give 
%the density line integrals over each of the $N_L$ chords $C^i$, $i=1, ...N_L$:
%$$
%N^i_{e,obs} \approx \int_{C^i} N_e(\psiN) dZ^i.
%$$
%
%Polarimetric measurements as they are considered in all former equilibrium reconstruction (e.g. 
%\cite{Hofmann:1988,Blum1997,Blum2012}) studies 
%give the Faraday rotation of the angle of the infrared radiation crossing the section of the plasma along the 
%different chords
%\begin{equation}
%\alpha^i_{obs} \approx \ds\frac{1}{2} \int_{C^i} \Omega_3 dZ^i.
%\label{eqn:polarClassic}
%\end{equation}
%
%As detailed in \cite{Segre1999} this is an approximation to one component of 
%the Stokes vector $\Vs(Z_1^i)$ only valid 
%for small Faraday and Cotton-Mouton effects. 
%On the contrary in this study we consider that 
%polarimetric measurements are given by the full Stokes vector at the 
%$Z_1^i$ coordinate on each chord $C^i$
%$$
%\Vs^i_{obs} \approx \Vs(Z_1^i).
%$$
%Indeed it is stated in \cite[Section 2]{Orsitto2011} and \cite[Section 8]{Segre1999} that the components of the Stokes 
%vector are related directly to quantities measured by the polarimetric system.

At this point we have defined a direct model given by the equilibrium 
equation (\ref{eqn:var}), 
%and Stokes equation (\ref{eqn:Stokes}) on every chord, 
control variables $\Acal$ and $\Bcal$,
% and $N_e$, 
and measurements to which are attached experimental errors 
represented by the standard deviations $\sigma$s in Eq. (\ref{eqn:Jobs}) below. 
The identification problem can now be formulated as a constrained minimization problem for 
the following cost function: 
%($\lbrace \Vs \rbrace$ denotes the vector $( \Vs^1, ..., \Vs^{N_L} )$ of Stokes vectors for all chords):

\begin{equation}
J(\psi,\Acal,\Bcal) :=  J_{obs}(\psi) + R(\Acal)+ R(\Bcal),
\label{eqn:Jcontinuous}
\end{equation}

where the least-square misfit term is
\begin{equation}
\begin{array}{l}
J_{obs}(\psi) :=  
 \ds \sum_{i=1}^{N_B} \frac{1}{2\sigma_{Bi}^2}(( \VB_p(r_i,z_i) \cdot \Vd_i ) 
- B^i_{p,obs})^2  
 + \ds \sum_{i=1}^{N_F} \frac{1}{2\sigma_{Fi}^2}(F(\psi)_i - F^i_{obs})^2 \\
\end{array}
\label{eqn:Jobs}
\end{equation}
and the regularization term $R(\Acal)$ is defined as
\begin{equation}
R(\Acal):= \ds \frac{\eps_{\Acal}}{2} \int_0^1 (\Acal''(x))^2 dx + \frac{\alpha_{\Acal}}{2} |\Acal(1)|^2.
\end{equation}
The $R(\Bcal)$ term is defined similarly.
Parameters $\eps$ enable to tune the smoothness of the identified functions 
whereas parameters $\alpha$ tune the penalization to zero of their value 
on the plasma boundary.
%\begin{equation}
%\begin{array}{l}
%R(\Acal,\Bcal) :=  
% \ds \frac{\varepsilon_A}{2} \ds \int_0^1 [\Acal''(x)]^2 dx  
%  + \ds \frac{\varepsilon_B}{2} \ds \int_0^1 [\Bcal''(x)]^2 dx 
%\end{array}
%\label{eqn:Jreg}
%\end{equation}
The inverse problem consists in the minimization of $J$ (\ref{eqn:Jcontinuous}) 
under the constraint of the model equation (\ref{eqn:var}).

\section{Numerical methods}
\label{sec:ip-discretization}
\subsection{Discretization of the direct model}

Equilibrium equation 
(\ref{eqn:var}) 
%(\ref{eqn:EquiModelConstraint}) 
is discretized using a P1 finite element 
method based on a triangular mesh. 
%\cite{Blum1981,Albanese1986,Heumann2015}. 
%The non-linearities are solved using Newton method. 
From now on let us also assume that functions $\Acal$ and $\Bcal$ are decomposed 
in a basis of functions ${\phi_i}$ 
defined on $[0,1]$. We use cubic spline functions in this work and
\begin{equation}
\begin{aligned}
\Acal(x)=\ds \sum_{i=1}^{N_{\Acal}} u_{\Acal i} \phi_i(x),\quad &  
\Bcal(x)=\ds \sum_{i=1}^{N_{\Bcal}} u_{\Bcal i} \phi_i(x). 
\end{aligned}
\end{equation}
Let us denote $\Vu=(\Vu_\Acal,\Vu_\Bcal)$ of size $N_u$ the vector of degrees of freedom 
of $\Acal$ and $\Bcal$ in the decomposition basis.
%\begin{equation}
%  \begin{aligned}
%  \Jsf_{\mathrm p}(\Vu)(\psi ,\xi) &:=
%  \int_{\Dp(\psi)}
% \lambda \left ( \ds \frac{r}{r_0} \ds \sum_{i=1}^{N} u_{Ai} \phi_i(\psiN) 
%+ \frac{r_0}{r} \ds \sum_{i=1}^{N} u_{Bi} \phi_i(\psiN)  \right) \xi \, dr dz
%\end{aligned}
% %\label{eq:Asf}
%\end{equation}
Classically approximating $\psi$ by $\psi_h=\ds \sum_{i=1}^{N} \psi_i \lambda_i(r,z)$ 
on the finite element approximation space 
%denoted $V_h$ 
as well as the operators of 
(\ref{eqn:var}) 
%(\ref{eqn:EquiModelConstraint}) 
and 
taking all basis elements $\lambda_i$ as test functions leads to 
the following non-linear system of $N$ equations:
\begin{equation}
(\Amat+\Cmat)\Vpsi - \VJ_p(\Vpsi, \Vu) - \Lmat \VI = 0
\label{eqn:EquiMat}
\end{equation}
where $\Vpsi$ denotes 
the vector of finite element coefficients $\lbrace \psi_i \rbrace_{i=1}^{N}$ 
and other notations are obvious.

\subsection{The discrete identification problem}

Using the discrete variables of the preceding section, cost function (\ref{eqn:Jcontinuous}) 
can also be discretized leading to the following expression 
\begin{equation}
\begin{array}{l}
J(\Vpsi,\Vu) :=  
 \ds \frac{1}{2} ||  \Hmat \Vpsi - \Vm) ||^2 + \ds \frac{1}{2} || \Rmat \Vu ||^2.
\end{array}
\label{eqn:Jdiscrete}
\end{equation} 
%\begin{equation}
%\begin{array}{l}
%J(\Vpsi,\lbrace \VS \rbrace, \Vu, \Vv) =  
% \ds \frac{1}{2} || \Dmat^{1/2} ( \Hmat \Vpsi - \Vm) ||^2 \\[10pt]
%+ \ds \frac{1}{2} || \Dmat^{1/2}_{N} ( \Wmat(\Vpsi)\Vv - \Vn_{obs}) ||^2 
% + \ds \sum_{i=1}^{N_L} \frac{1}{2}||\Dmat^{1/2}_{Si}(\Emat \VS^i - \Vs^i_{obs})||^2 \\[10pt]
% + \ds \frac{\varepsilon}{2} || \Rmat_u \Vu ||^2 + \ds \frac{\varepsilon_n}{2} || \Rmat_v \Vv ||^2 
%\end{array}
%\label{eqn:Jdiscrete}
%\end{equation} 
 %\ds \frac{1}{2} || \Dmat^{1/2}_{B} ( \Hmat_B \Vpsi - \VB_{p,obs}) ||^2 
 %+ \ds \frac{1}{2} || \Dmat^{1/2}_{F} ( \Hmat_F \Vpsi - \Vpsi_{obs}) ||^2 \\[10pt]
%Here $\lbrace \VS \rbrace$ is the vector $(\VS^1, ...,\VS^{N_L})$. 
In order to lighten 
notations the $\ds \frac{1}{\sigma}$ terms have been dropped  
and are assumed to be included in the observation operator $\Hmat$ and in the measurements $\Vm$.
%the $\Dmat^{1/2}$ matrices are diagonal matrices containing the observation errors 
%$\ds \frac{1}{\sigma}$, 
%The linear observation operator $\Hmat$ maps the finite element 
%approximation $\Vpsi$ to the equivalent of magnetic measurements $\Vm$. 
%The non-linear observation operator $\Wmat(\Vpsi)\Vv$ represents the numerical quadrature of the electron 
%density over the different chords. 
%This term is linear in $\Vv$ but not in $\Vpsi$. 
%The observation operator for Stokes vectors is given by matrix $\Emat$ 
%such that $\Emat \VS^i = \Vs^{i,N^i}$ is the Stokes vector at the observation point. 
The last term involving matrice $\Rmat$ is the discretization of the regularization terms 
in which we have gathered the contributions from functions $\Acal$ and $\Bcal$. 
%and chosen $\varepsilon_A= \varepsilon_B =\varepsilon_{N_e}$.

The discrete identification problem can now be stated as 
\begin{equation}
\underset{\Vpsi, \Vu}{\min} J(\Vpsi, \Vu)
\label{eqn:opt1}
\end{equation}
subject to the constraint of the non-linear model
\begin{equation}
(\Amat+\Cmat)\Vpsi - \VJ_p(\Vpsi, \Vu) - \Lmat \VI = 0
\label{eqn:opt2}
\end{equation}

This problem is solved thanks to a quasi-SQP algorithm with reduced Hessian (QSQP).
SQP methods are well documented \cite{Nocedal2006,Hinze2009} and for fusion application we refer to 
\cite[Appendix A]{Blum:2019} and \cite{ACL.B.Faugeras.20.2, ACL.B.Faugeras.17.2}. 
An SQP method can be seen as a Newton method to solve the non-linear system given 
by the fisrt order optimality condition for the Lagrangian of the PDE-constrained optimization problem.

The QSQP method we use is the following 2 steps iterative algorithm:

\label{sqp}
\begin{enumerate}
\item control variable update step
\begin{equation}
\label{eqn:ustep}
\Mmat(\Vu^{n+1}-\Vu^n)=-\Vh
\end{equation}
\item state variable update step
\begin{equation}
\label{eqn:ystep}
\Vpsi^{n+1}=\Vpsi^n +\delta \Vpsi +\Smat  (\Vu^{n+1}-\Vu^n)
\end{equation}
\end{enumerate}
where
\begin{equation}
\label{eqn:dy}
\delta \Vpsi= -[(\Amat+\Cmat)\Vpsi^n - D_{\Vpsi}\VJ_p(\Vpsi^n, \Vu^n) ]^{-1} [(\Amat+\Cmat)\Vpsi^n - \VJ_p(\Vpsi^n, \Vu^n) - \Lmat \VI],
\end{equation}
\begin{equation}
\label{eqn:S}
\Smat= -[(\Amat+\Cmat)\Vpsi^n - D_{\Vpsi}\VJ_p(\Vpsi^n, \Vu^n) ]^{-1} [- D_{\Vu}\VJ_p(\Vpsi^n, \Vu^n)],
\end{equation}

\begin{equation}
\label{eqn:M}
\Mmat=\Rmat^T\Rmat +\Smat^T \Hmat^T \Hmat \Smat
\end{equation}
%:=\Zmat^T_k 
%\begin{bmatrix}[cc]
%J_{\Vy \Vy}(\Vy^k,\Vu^k) &  J_{\Vy \Vu}(\Vy^k,\Vu^k)\\
%J_{\Vu \Vy}(\Vy^k,\Vu^k) &  J_{\Vu \Vu}(\Vy^k,\Vu^k)\\
%\end{bmatrix}
%\Zmat_k
 
and
\begin{equation}
\label{eqn:m}
\Vh= \Rmat^T \Rmat \Vu^n + \Smat^T \Hmat^T ( \Hmat \Vpsi^n- \Vm) + 
  \Smat^T \Hmat^T \Hmat \delta \Vpsi
\end{equation}

At each iteration this algorithm demands the resolution of $N_u+1$ linear systems (\ref{eqn:dy})-(\ref{eqn:S}) 
of size $N$ involving the same matrix with different righ-hand sides which can be done very efficiently 
and of one smaller linear system of size $N_u$ in (\ref{eqn:ustep}).

The performance of the QSQP method used for the resolution 
of the identification problem relies on the accuracy of the derivative terms 
$D_\psi \VJ_p$,  $D_\Vu \VJ_p$. 
In this work we have implemented the exact derivatives of the fully discretized operators. 
This essential but very technical 
work is not further detailed here and we refer to \cite[section 3.2 and 3.3]{Heumann2015} 
for details.

%A first initialization step is used to set $\lambda$ in the current density term given the total plasma current $I_p$ 
%which is either given with magnetic measurements or comes out from the toroidal harmonics procedure. 
%Given initial guess $\Vpsi^0$, $\Vu_{\Acal}^0$ and $\Vu_{\Bcal}^0$, $\lambda$ is chosen to satisfy
%\begin{equation}
%I_p-\lambda J_p(\Vpsi^0, \Vu_{\Acal}^0, \Vu_{\Bcal}^0)=0 
%\end{equation}

\section{Numerical experiment}
\label{sec:num}

The numerical methods presented in the previous sections are implemented in the code NICE 
\cite{ACL.B.Faugeras.20.2} with which the following numerical experiment is conducted. 
ITER machine description providing the points defining the limiter contour, the poloidal field coils and a description of the 
magnetic sensors (196 poloidal magnetic field probes and 22 differential flux loops) are read 
from the {\it wall}, {\it pf$\_$active} and {\it magnetics} Interface Data Structure (IDS) 
from the ITER Integrated Modelling and Analysis Suite (IMAS) \cite{IMAS2015}. 
From this machine description NICE builds the triangular mesh used with the finite elements computations. 

We generate a reference equilibrium by solving the so-called inverse static equilibrium problem 
\cite{Heumann2015,ACL.B.Faugeras.20.2} that is to say finding the currents in the coils giving a desired prescribed plasma boundary. 
For this reference computation the unknown function are given analytically as 
$\Acal(x)=(1-x^{1.5})^{0.9}$ and $\Bcal(x)=(1-x^{0.9})^{1.5}$. 
The scaling factor $\lambda$ in the current density (\ref{eqn:density}) 
is computed such that the total plasma current is $I_p=12$ $[MA]$. 
The vacuum toroidal field is $B_0=5.3$ $[T]$ and $r_0=6.2$ $[m]$. 
Synthetic magnetic measurements are computed from this reference equilibrium. 

Then in a second step these measurements are plugged in cost function (\ref{eqn:Jdiscrete}) 
and the optimization problem is solved using the 
iterative QSQP algorithm presented above. 
The initial guess for this resolution consists in a given circular plasma domain in which 
the flux $\psi$ is a constant and outside of which it is $0$, as well as 
affine functions $\Acal(x)=\Bcal(x)=1-x$. 
%A first Newton iteration for the direct equilibrium 
%problem with fixed $I_p$ \cite{Heumann2015} 
%is performed in order to compute a first $\psi$ map and to give a value to the scaling factor $\lambda$. 
%Thereon this factor is kept fixed and for what concerns the plasma current density only 
%$\Vu$ representing the functions to be identified evolves during the iterations. 
Convergence is assumed when the relative residue for the vector of unknowns $\VX=(\Vpsi,\Vu)$ satisfies 
${||\VX^{k+1}-\VX^{k}||}/{||\VX^{k}||}<10^{-12}$. 
%The convergence history for this test case is shown on table \ref{tab:convergence}.

The finite element mesh is composed of $N = 11856$ nodes among which $11772$ correspond 
to free values of $\psi$ (the remaining correspond to the imposed boundary condition $\psi=0$ on the axis $r=0$). 
%The discretization step on each chord is chosen to 
%be $h=0.05$ $[m]$ giving a vector $\lbrace \VS \rbrace$ of size $3\times2022$. 
Each function to be identified is decomposed 
in $11$ cubic splines defined on $[0,1]$ with knots at $0, 0.1, ...,1$. 
Therefore $\VX$ is a vector of size $11772+2\times11=11794$.% for type M experiments, 
%$2\times30270+3\times11=60573$ for type MF and $2\times(30270+3\times2022)+3\times11=72705$ for type MS experiments.

The regularization parameters $\varepsilon$ are tuned to their lowest value, typically $10^{-2}$, 
avoiding oscillations in the reconstructed profiles or non convergence of the code. The penalization parameters  
are set to  $\alpha = 10^{12}$.

\begin{figure}
\includegraphics[width=0.65\textwidth]{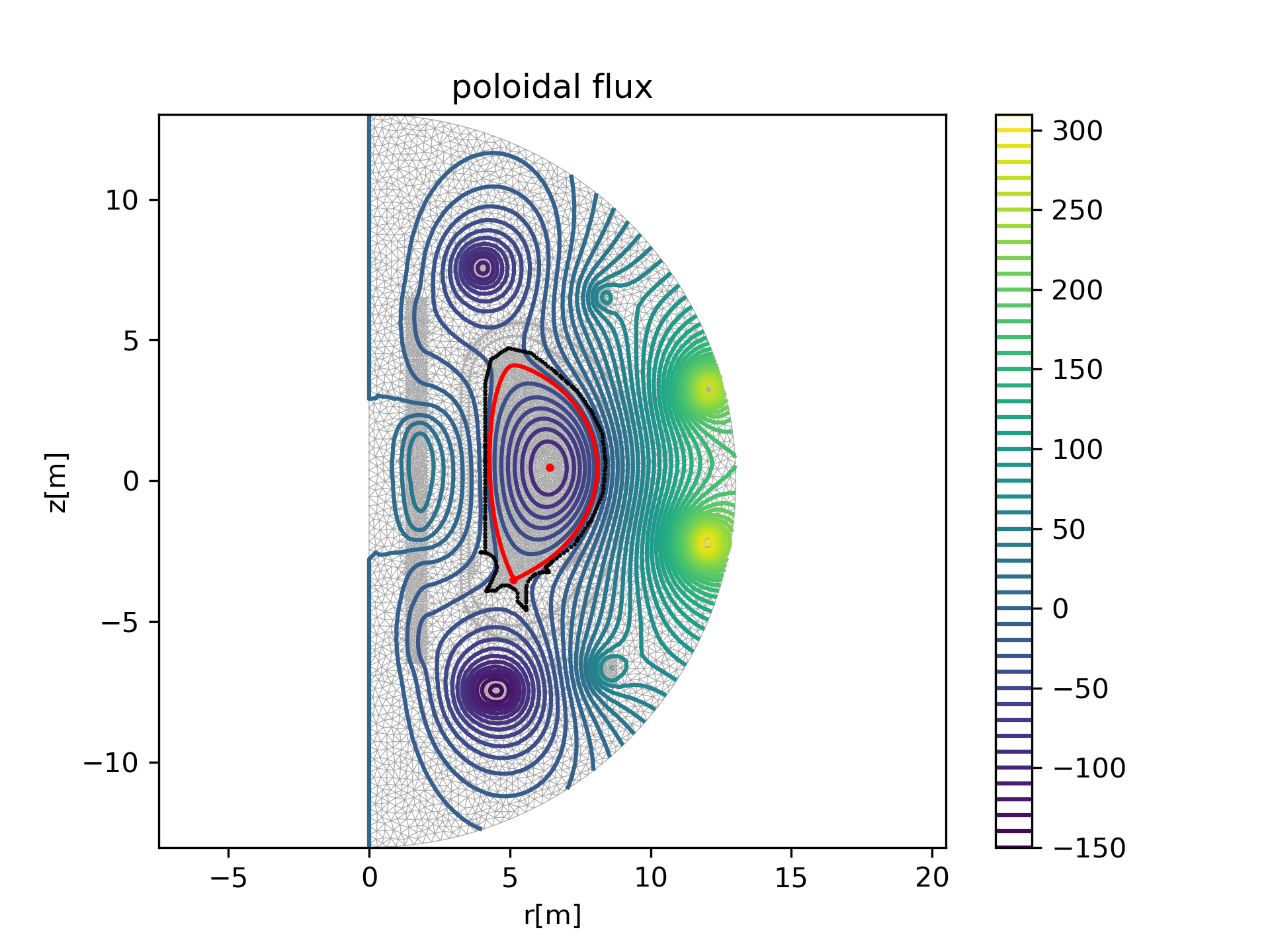}
\caption{Reconstructed poloidal flux map ($\psi$ in [Wb]). The plasma boundary with an X-point and the magnetic axis are shown in red.}
\label{fig:fluxmap}
\end{figure}

Starting from the initial guess described above, which is far from the reference solution we want to recover, 
the algorithm needs 16 iterations to converge to a relative residue of $4 \times 10^{-14}$. In this test configuration 
one iteration takes about $800$ ms on a laptop with an Intel CPU at $2.90$ GHz. This computation time is highly dependent 
on the mesh size and is limited by the performance of the linear solver.
The reconstructed flux map is shown on Figure \ref{fig:fluxmap}. 
The plasma boundary is perfectly recovered. The fit to measurements is excellent with a root mean square error of 
$10^{-4}$ on magnetic probes and flux loops.

\begin{figure}
\includegraphics[width=0.65\textwidth]{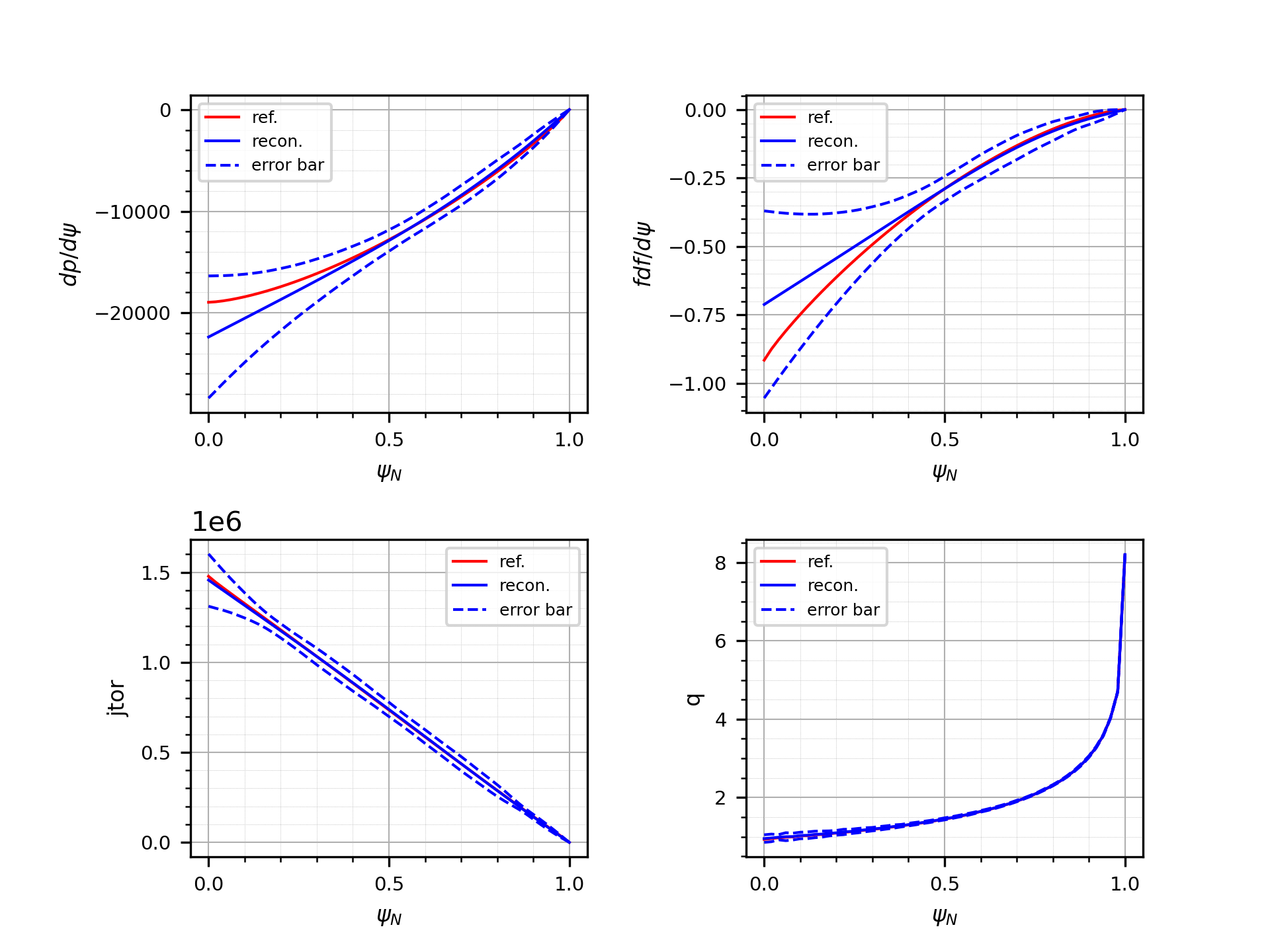}\\
\caption{Top row: reference and reconstructed $p'$ and $ff'$ profiles as function of the normalized flux $\psiN$ with computed error bars. Bottom row: same for the average toroidal current density $jtor$ and safety factor $q$ profiles.}
\label{fig:profiles}
\end{figure}

Figure \ref{fig:profiles} shows the reference and identified $p'$ and $ff'$ profiles. A typical feature of the equilibrium reconstruction 
inverse problem using only magnetic measurements appears: the computed error bars on the reconstructed profiles 
increase when approaching the magnetic axis. 
However even if the $p'$ and $ff'$ profiles are not perfectly identified for low $\psiN$ values, 
the flux surface averaged current density profile $jtor=<j_\phi/r>/<1/r>$ 
and the safety factor $q$ are very well recovered.

\section{Conclusion}

The equilibrium code NICE is ready to use ITER data. A first equilibrium reconstruction numerical exercice 
using synthetic magnetic measurements has been successfully conducted. The code is fully IMAS compatible, can read and write IDS. 
It is also ready to use other measurements such as pressure measurements, motional Stark effect (MSE) as well as interferometry and polarimetry.  
The polarimetry Stokes vector modelization can also be used \cite{ACL.B.Faugeras.18.2}. These internal measurements will most likely be needed to improve the accuracy 
of the reconstruction of the $p'$ and $ff'$ profiles for ITER discharges. 
The code NICE is already routinely used at WEST and has been tested on different tokamaks and validated against 
other codes within the EUROfusion program \cite{ACTI.B.Faugeras.21.6,COM.B.Faugeras.20.5,ACTI.B.Faugeras.19.5,ACL.B.Faugeras.21.3,ACL.B.Faugeras.19.3}. 
NICE is a robust code which exhibits excellent convergence properties thanks to the use of Newton and SQP methods. 
The code can be used for equilibrium reconstruction but also for direct static or evolutive simulations. 
It can also be used to solve the inverse problems consisting in finding the currents or voltages in the poloidal field coils 
which enable to have a desired plasma shape \cite{ACL.B.Faugeras.20.2}. Finally the code NICE can be used with high order 
finite elements providing a smooth representation of the magnetic flux and field in the plasma \cite{ACL.B.Faugeras.21.1,ACTI.B.Faugeras.21.4}.

\clearpage
\newpage

\section*{Acknowledgment}
We would like to thank Masanari Hosokawa, Simon Pinches and Mireille Schneider 
for their help in providing the latest ITER machine description. 
We would also like to thank the anonymous reviewer for the numerous remarks and constructive criticism
from which the paper has benefitted.% significantly.

%\section*{References}

\bibliographystyle{plain}

\bibliography{biblio}

\end{document}